# Tuning the $S = 1/2$ square-lattice antiferromagnet $Sr_2Cu(Te_{1-x}W_x)O_6$ from Néel order to quantum disorder to columnar order

O. Mustonen,[1] S. Vasala,[2] K. P. Schmidt,[3] E. Sadrollahi,[3] H. C. Walker,[4] I. Terasaki,[5] F. J. Litterst,[2,3] E. Baggio-Saitovitch,[2] M. Karppinen[1*]

[1]*Department of Chemistry and Materials Science, Aalto University, FI-00076 Espoo, Finland*

[2]*Centro Brasileiro de Pesquisas Físicas (CBPF), Rua Dr Xavier Sigaud 150, Urca, Rio de Janeiro, 22290-180, Brazil*

[3]*Institut für Physik der Kondensierten Materie, Technische Universität Braunschweig, 38110 Braunschweig, Germany*

[4]*ISIS Neutron and Muon Source, Rutherford Appleton Laboratory, Chilton, Didcot, OX11 0QX, United Kingdom*

[5]*Department of Physics, Nagoya University, Nagoya 464-8602, Japan*

\* Corresponding author

The spin-1/2 square-lattice Heisenberg model is predicted to have a quantum disordered ground state when magnetic frustration is maximized by competing nearest-neighbor $J_1$ and next-nearest-neighbor $J_2$ interactions ($J_2/J_1 \approx 0.5$). The double perovskites $Sr_2CuTeO_6$ and $Sr_2CuWO_6$ are isostructural spin-1/2 square-lattice antiferromagnets with Néel ($J_1$ dominates) and columnar ($J_2$ dominates) magnetic order, respectively. Here we characterize the full isostructural solid solution series $Sr_2Cu(Te_{1-x}W_x)O_6$ ($0 \leq x \leq 1$) tunable from Néel order to quantum disorder to columnar order. A spin-liquid-like ground state was previously observed for the $x = 0.5$ phase, but we show that the magnetic order is suppressed below 1.5 K in a much wider region of $x \approx 0.1$-0.6. This coincides with significant $T$-linear terms in the low-temperature specific heat. However, density functional theory calculations predict most of the materials are not in the highly frustrated $J_2/J_1 \approx 0.5$ region square-lattice Heisenberg model. Thus, a combination of both magnetic frustration and quenched disorder is the likely origin of the spin-liquid-like state in $x = 0.5$.

# I. INTRODUCTION

Square-lattice antiferromagnets (AFMs) can be described using the Heisenberg $J_1$-$J_2$ model with two interactions: nearest neighbor $J_1$ (NN; side of the square) and next-nearest neighbor $J_2$ (NNN; diagonal of the square). The phase diagram of the model is shown in Fig. 1a. The classical ground states for the $J_1$-$J_2$ model are ferromagnetic order, Néel AFM order and columnar AFM order[1]. Dominating antiferromagnetic $J_1$ ($J_2/J_1 \ll 0.5$) leads to Néel order and dominating $J_2$ ($J_2/J_1 \gg 0.5$) to columnar order. Anderson[2] first proposed that by frustrating the Néel order in a $S = 1/2$ square-lattice AFM, i.e. by introducing an antiferromagnetic $J_2$ interaction, a quantum spin liquid (QSL) state might emerge. Quantum spin liquids are highly entangled quantum states with exotic excitations, in which spins remain dynamic and do not order even at absolute zero[3–5]. The QSL state on the square lattice is predicted to occur in a narrow parameter range between $J_2/J_1 \approx 0.4$-$0.6$, where the magnetic frustration due to competing antiferromagnetic $J_1$ and $J_2$ interactions is maximized[6–8]. A limited number of model compounds that realize the $S = 1/2$ square-lattice Heisenberg model are known[9–19], but none of them are in the spin liquid region of the phase diagram (Fig. 1a).

Recently, $B$-site ordered $A_2B'B''O_6$ double perovskites $Sr_2CuTeO_6$ and $Sr_2CuWO_6$ were shown to be near-ideal realizations of the $S = 1/2$ square-lattice $J_1$-$J_2$ model with highly two-dimensional magnetic interactions[20–23]. These two compounds are unique among the known $J_1$-$J_2$ model compounds, because they are isostructural yet on the opposite sides of the phase diagram: the magnetic ordering is Néel type in $Sr_2CuTeO_6$ ($J_2/J_1 = 0.03$) and columnar in $Sr_2CuWO_6$ ($J_2/J_1 = 7.92$)[22–25]. The crystal structure is tetragonal with a rocksalt ordering of $B'$ ($Cu^{2+}$) and $B''$ ($Te^{6+}/W^{6+}$) sites[26–28], see Fig. 1b. The copper cations form a square in the $ab$ plane, and the diamagnetic $Te^{6+}/W^{6+}$ cations is located in the middle of the square, as shown in Fig. 1c. The diamagnetic $B''$ cation controls the extended superexchange pathways between the copper cations: $W^{6+}$ $5d^0$ hybridizes strongly with O $2p$ and allows 180° Cu-O-W-O-Cu ($J_2$) superexchange, which is not possible with the $4d^{10}$ $Te^{6+}$ cations which favor $J_1$[22,29,30]. As a result, the two compounds are in different regions of the $J_1$-$J_2$ phase diagram despite being isostructural.

The solid solution $Sr_2Cu(Te_{1-x}W_x)O_6$ is a unique system for studying frustrated square-lattice antiferromagnetism as it can be tuned from Néel ($x = 0$) to columnar order ($x = 1$) by varying the composition. Additionally, the similar size of $Te^{6+}$ and $W^{6+}$ cations means that little change is expected in the crystal structure[31]. Very recently, we showed that Sr$_2$Cu(Te$_{0.5}$W$_{0.5}$)O$_6$ ($x = 0.5$) has a spin-liquid-like ground state[32], which exhibits many of the properties of the QSL state: magnetism was found to be entirely dynamic down to 19 mK and a plateau was observed in the low-temperature muon spin relaxation rate. Moreover, the magnetic specific heat showed a strong $T$-linear relationship at low temperatures indicating gapless excitations[32].

The origin of the spin-liquid-like ground state in Sr$_2$Cu(Te$_{0.5}$W$_{0.5}$)O$_6$ ($x = 0.5$) is not known. The QSL state expected from the $J_1$-$J_2$ model requires a good match of the magnetic interactions $J_1$ and $J_2$[6–8]. Thus, it is expected to be stable only in a narrow composition range in Sr$_2$Cu(Te$_{1-x}$W$_x$)O$_6$. To illustrate this point we can, as a first approximation, interpolate the exchange interactions in the Sr$_2$Cu(Te$_{1-x}$W$_x$)O$_6$ solid solution from the end members Sr$_2$CuTeO$_6$ ($J_1$ = -7.18 meV, $J_2$ = -0.21 meV) and Sr$_2$CuWO$_6$ ($J_1$ = -1.2, $J_2$ = -9.5 meV) assuming a linear dependence on $x$[22,23]. In such a case, the highly frustrated region with $J_2/J_1 \approx 0.4$-$0.6$ would correspond to a rather narrow composition range of $x \approx 0.23$-$0.33$. However, the Sr$_2$Cu(Te$_{1-x}$W$_x$)O$_6$ solid solution system has significant quenched disorder in the magnetic interactions that is not included in the $J_1$-$J_2$ model. This is due to Te$^{6+}$/W$^{6+}$ disorder on the $B''$ site in the middle each square of Cu$^{2+}$ ions (Fig. 1c), which affects whether $J_1$ or $J_2$ is dominant locally. This quenched disorder could help to stabilize a QSL[33–36] or a QSL-like random-singlet state[37–41].

Here we investigate the properties of the full solid solution series Sr$_2$Cu(Te$_{1-x}$W$_x$)O$_6$ ($0 \leq x \leq 1$). Our magnetic susceptibility measurements suggest a maximum in frustration near $x \approx 0.4$-$0.5$, whereas low-temperature specific heat measurements show a $T$-linear term typical of QSLs or spin glasses in a wide composition range of $x \approx 0.1$-$0.7$. Similarly, muon spin relaxation and rotation measurements reveal that magnetic order is significantly suppressed not only in $x = 0.5$ but for $x \approx 0.1$-$0.6$. Whether the QSL-like state of $x = 0.5$ occurs for this entire composition range or another ground state such as spin glass forms is not known at this time. Density functional theory (DFT) calculations suggest a crossover from Néel ($J_2/J_1 < 0.5$) to columnar ($J_2/J_1 > 0.5$) region occurs already at $x \approx 0.2$ placing most samples in the columnar region. Thus, magnetic frustration from competing interactions on its own does not explain the suppression of magnetic order in this system: a combination of both magnetic frustration and quenched disorder is required.

## II. EXPERIMENTAL DETAILS

The solid solution series $Sr_2Cu(Te_{1-x}W_x)O_6$ ($0 \leq x \leq 1$) was synthesized by conventional solid state synthesis. Stoichiometric amounts of $SrCO_3$, $CuO$, $TeO_2$ and $WO_3$ were ground in an agate mortar with ethanol. The precursor mixture was calcined at 900 °C for 12 h, pelletized and repeatedly fired at 1050 °C in air for a total of 72 h with intermittent grindings. The synthesis temperature was of special importance on the Te-rich side ($x < 0.5$), as too high a temperature was found to result in the formation of an unknown impurity phase with main reflections at $2\theta$ (Cu $K_{\alpha 1}$) 30.56° and 30.76°. At the same time, the synthesis temperature needed to be high enough to form a well crystallized double perovskite phase. We found no positive effect from using excess $TeO_2$ in contrast to an earlier report on $Sr_2CuTeO_6$[27].

Phase purity and crystal structure of the samples were analyzed by powder x-ray diffraction (Panalytical X'Pert Pro MPD, Cu $K_{\alpha 1}$ radiation). Rietveld refinements were carried out using FULLPROF software[42]. Line broadening analysis was performed as described in ref. [43]. Instrumental broadening was determined with a $LaB_6$ standard (NIST SRM 660b). The crystal structures were visualized using VESTA 3[44].

Magnetic properties were measured with a Quantum Design MPMS XL magnetometer. The samples were measured in gelatin capsules placed inside plastic straws. Magnetic susceptibility was measured from 5 to 300 K in an applied field of 1 T. Specific heat was measured with a Quantum Design PPMS instrument. The data were collected between from 2 to 150 K using a thermal relaxation method.

Muon spin rotation and relaxation (µSR) for polycrystalline powders with compositions $x = 0.1, 0.2, 0.3, 0.4, 0.6, 0.7, 0.8$, and $0.9$ were measured at the Dolly and GPS installations of the Swiss Muon Source at Paul Scherrer Institut, Switzerland, using 100% spin-polarized positive muons. Sample temperatures were varied down to 1.5 K. The $x = 0.5$ sample was previously measured down to 19 mK at the LTF installation[32]. Measurements were performed in zero-field (ZF) and weak transverse-field (wTF) mode (5 mT applied perpendicular to initial muon spin direction).

Density functional theory calculations were used to evaluate the relative stabilities of different magnetic orderings in the full composition range $0 \leq x \leq 1$ in steps of 1/8 (0.125) in $x$. We have previously shown that our computational approach works well for $Sr_2CuWO_6$ and gives a good estimate of the exchange coupling constants[23]. The calculations were carried out with the full-potential linearized augment plane-wave ELK code[45] using the the generalized gradient approximation functionals by Perdew, Burke and Ernzerhof[46]. Electron correlation effects of the 3$d$ copper orbitals were included with a DFT+$U$ approach using the fully-localized limit double counting correction[47]. We used a Hubbard $U$ value of $U = 8$ eV typical of copper[47–49] (with Hund's rule coupling $J = 0.9$ eV corresponding to $U_{eff} = 7.1$ eV), which was also the optimal value for $Sr_2CuWO_6$[23]. The calculations were carried out using 2 x 2 x 1 supercells with ferromagnetic, Néel or columnar antiferromagnetic order[23]. The experimental crystal structures were used without lattice relaxation. Only one supercell was used for each composition with an equal number of Te and W nearest neighbors for $x = 0.5$ (Supplemental Material[50]). We used a $k$ point grid of 4 x 4 x 6 and a plane-wave cut-off of $|G + k|_{max} = 8/R_{MT}$ a.u.$^{-1}$, where $R_{MT}$ is the average muffin-tin radii.

## III. RESULTS

### A. Crystal structure

We were able to synthesize the full solid solution series $Sr_2Cu(Te_{1-x}W_x)O_6$ ($0 \leq x \leq 1$) using a conventional solid state reaction method. Powder x-ray diffraction revealed the samples to be of high quality with trace quantities (less than 1%) of a $SrWO_4$ impurity. Rietveld refinement revealed that all the samples crystallize in the $B$-site ordered double perovskite structure of the parent phases (Supplemental Material[50]). The space group is $I4/m$ in all cases. Since $Te^{6+}$ and $W^{6+}$ cations have very similar sizes[31], we see only small changes in the lattice parameters. The main changes are seen in the $c$ parameter, which decreases with increasing $x$. However, small changes are also observed in the $a$ parameter: it stays relatively constant below $x = 0.5$ and then starts decrease very slightly with increasing $x$. The unit cell volume changes linearly and follows Vegard's law, which shows that we have prepared a true solid solution. The changes in atomic positions are minor. The $B$' ($Cu^{2+}$) and $B$'' ($Te^{6+}/W^{6+}$) cations appear fully ordered, but there is no sign of $Te^{6+}$ and $W^{6+}$ ordering on the $B$'' site consistent with our previous report on $x = 0.5$[32]. The structural disorder on the $B$'' cation site, which determines whether $J_1$ or $J_2$ dominates, inevitably leads to some quenched disorder in the magnetic interactions between the $Cu^{2+}$ sites.

### B. Magnetic properties

Magnetic susceptibility in all samples features a broad maximum (Fig. 2a). While the end members $Sr_2CuWO_6$ and $Sr_2CuTeO_6$ order antiferromagnetically, the Néel temperatures cannot be determined from the susceptibility as no cusp is observed[21,24]. The broad maximum is expected behavior for a $J_1$-$J_2$ model square-lattice antiferromagnet. It can be characterized by its position $T_{max}$ and height $\chi_{max}$. Theoretical studies on magnetic susceptibility in the $J_1$-$J_2$ model as a function of $J_2/J_1$ indicate that $T_{max}$ should have a minimum at $J_2/J_1 = 0.5$ and a $\chi_{max}$ should have a maximum at the same $J_2/J_1$ ratio[51,52]. We see a similar effect in $Sr_2Cu(Te_{1-x}W_x)O_6$ (Fig. 2b): $T_{max}$ has a minimum at x ≈ 0.4-0.6 whereas $\chi_{max}$ has a maximum at $x ≈ 0.4$, see Table I. This suggests that the magnetic frustration is at its highest near $x ≈ 0.4$-$0.5$. As we have previously shown, $x = 0.5$ has a spin-liquid-like ground state[32]. The behaviors of $T_{max}$ and $\chi_{max}$ in the $Sr_2Cu(Te_{1-x}W_x)O_6$ series are clearly different from the analogous molybdenum-based series $Sr_2Cu(Mo_{1-x}W_x)O_6$, where both depend linearly on composition[53]. This difference is expected, since $Sr_2CuTeO_6$ has Néel order[22,24] but both $Sr_2CuWO_6$ and $Sr_2CuMoO_6$ have columnar order[21,23,25] resulting in little magnetic frustration in $Sr_2Cu(Mo_{1-x}W_x)O_6$. Additionally, low-temperature Curie tails are observed for the samples $x = 0.1$-$0.7$.

In order to evaluate the overall strength of magnetic interactions in the materials, the Curie-Weiss temperatures $\Theta_{cw}$ were obtained by fitting the magnetic susceptibilities to the Curie-Weiss law $\chi = C/(T-\Theta_{cw})$. The data were fitted in the temperature range 250–300 K. As shown in Table I, the Curie-Weiss temperatures $\Theta_{cw}$ for all samples are negative indicating antiferromagnetic interactions. The overall strengths of magnetic interactions in the Te-rich side ($x < 0.5$) are very similar to each other but slightly weaker than in $Sr_2CuTeO_6$ as indicated by the lower $\Theta_{cw} ≈ -60$ K. In the W-rich side the interactions become stronger with increasing $x$ with the strongest AFM interactions in $Sr_2CuWO_6$ ($\Theta_{cw} = -165$ K). This trend corresponds to the minor changes observed in the lattice parameter $a$, which is directly related to the $Cu^{2+}$-$Cu^{2+}$ distance in the $ab$ plane. The effective paramagnetic moments obtained from the fits were similar for all samples and typical for $Cu^{2+}$ compounds.

### C. Specific heat

Specific heat of selected samples is shown in Fig. 3a. No lambda anomalies expected for long-range magnetic ordering at $T_N$ are observed in any of the samples. The main differences between the samples are in the low-temperature specific heat. In order to evaluate the temperature dependence we plot the reduced low-temperature specific heat as a function of $T^2$ (Fig. 3b). A strong $T$-linear $\gamma$ term in specific heat is expected for a QSL. This $T$-linear term is related to excitations of highly entangled spins. For $x = 0$, 0.9 and 1 the reduced specific heat approaches zero with decreasing temperature. This lack of a $T$-linear term is expected behavior in long-range ordered antiferromagnetic insulators. For the other samples ($x = 0.1$-$0.8$) we observe behavior indicating a $T$-linear term, but at very low temperatures some differences are observed. In $x = 0.7$ and 0.8 the reduced specific heat has an additional small downturn at very low temperatures, suggesting that the $T$-linear $\gamma$ term could be significantly smaller or even zero in these samples. In the $x = 0.1$-0.6 samples the reduced specific heat either remains linear or has a slight up-turn at very low temperatures.

The specific heat data below 10 K were fitted using the function $C_p = \gamma T + \beta_D T^3$, where $C_p$ is the specific heat, $\gamma$ is the electronic $T$-linear term and $\beta_D$ is the phononic term. We previously found very small $\gamma$ terms for $Sr_2CuTeO_6$ and $Sr_2CuWO_6$, but a significant one for $x = 0.5$, which is typical for gapless spin-liquid-like states[54] and spin glasses[55]. For the full series we find a significant $\gamma$ term at a very wide composition range of $x = 0.1$-0.7 (Fig. 3c). This result suggests that the spin-liquid-like state could be present in a wide composition range, although a spin glass ground state cannot be ruled out for compositions other than $x = 0.5$. The largest electronic contribution occurs at $x = 0.5$ in line with the magnetic susceptibility results.

### D. Muon spin rotation and relaxation

Zero-field (ZF) μSR on $Sr_2CuWO_6$ ($x = 1$) has been reported to exhibit spontaneous rotation signals revealing long-range magnetic order below $T_N = 24$ K with a single rotation frequency following a typical temperature dependency curve[21]. Our ZF spectra for $x = 0.9$, 0.8, and 0.7 are presented in Fig. 4a, b and c, where we plot the time dependent polarization $G_z(t)$ of the muon spins measured via the asymmetry of decay positron count rates $A(t)/A(0) = G_z(t)$ after muon implantation at $t = 0$. Similar to $x = 1$, these spectra also show spontaneous rotations indicative of long-range order. However, these are strongly damped and with the onsets at clearly lower temperatures, around 15 and 11 for $x = 0.9$ and 0.8, respectively. For $x = 0.7$ the oscillations are less clearly visible, and the onset of magnetic order around 7 K can better be traced from wTF experiments, as will be described below. In contrast to $x = 1$ where only a single rotation signal was sufficient to yield good fits to the data, we need three signals for $x = 0.9$, 0.8 and 0.7 with different temperature independent asymmetries $A_i$, temperature dependent frequencies $f_i$, and transverse and longitudinal damping factors $\lambda_{ti}$ and $\lambda_{li}$ ($i = 1, 2, 3$). While usually rotation signals present cosine shape, we had to use zeroth order Bessel functions $J_o$ for the best fits.

The total asymmetry is thus given by

$$A_{\text{tot}} = \sum_i A_i \cdot G_{zi} \cdot G_{\text{KT}i} \tag{1}$$

where

$$G_{zi} = \tfrac{2}{3} \exp(-\lambda_{ti} t) J_0(2\pi f_i t) + \tfrac{1}{3} \exp(-\lambda_{li} t) \tag{2}$$

represents the rotations and dampings caused by magnetic fields of electronic origin. While the damping $\lambda_{ti}$ of the oscillating signals is mainly caused by inhomogeneous static fields, the damping $\lambda_{li}$ of 1/3 of asymmetry (second term) is caused by dynamic spin relaxation. The local magnetic fields $B_i$ acting at the muon sites are related to the rotation frequencies $f_i$ via

$$B_i = (2\pi/\gamma_\mu) \cdot f_i \tag{3}$$

where $\gamma_\mu/2\pi = 135.5$ MHz/T is the muon gyromagnetic ratio. $G_{KTi}$ is a so-called static Kubo-Toyabe function representing the depolarization due to a distribution of static nuclear dipolar fields:

$$G_{KTi} = \tfrac{2}{3}(1 - \sigma^2 t^2)\exp\left(-\frac{\sigma^2 t^2}{2}\right) + \tfrac{1}{3} \tag{4}$$

where σ is the width of a Gaussian distribution of static fields. For $x = 0.9$, 0.8 and 0.7 we use σ = 0.07, 0.06 and 0.04 µs$^{-1}$, respectively, as estimated from damping functions obtained at high temperature where damping from electronic origin is negligible.

The observation of Bessel-shaped rotation signals is typical for a wide distribution of frequencies and is typically met in materials with incommensurately modulated spin structures.[56,57] The derived internal fields $B_i$ are corresponding to the maximum cut-off frequencies $f_i$ of the frequency distribution. In our present case, we use Bessel-shaped rotation signals simply as a heuristic approximation to a complex distribution. A more detailed description of analysis and discussion will be given elsewhere. For the present considerations, we restrict ourselves to a qualitative discussion of the µSR data. In Fig. 4d we compare the temperature dependent local fields $B_i$ obtained for $x = 0.9$ with those of $x = 1$ from ref. [21]. It is clear that the ordering temperature and local fields are reduced for $x = 0.9$ compared to $x = 1$. This is further seen in the local fields derived for $x = 0.8$ and 0.7 shown in the Supplemental Material[50]. For a rough estimate of the ordering temperatures $T_N$ we have extrapolated the $B_i(T)$ using the relation

$$B_i(T) = B_0 \left[1 - \left(\frac{T}{T_N}\right)^\alpha\right]^\beta \tag{5}$$

with α = 1.33, and β = 0.5 for $x = 0.7$ and 0.8, and β = 0.3 for $x = 0.9$. Especially for $x = 0.7$ and 0.8 the extrapolations give varying values for the different $B_i$ at one concentration indicating a range of ordering temperatures. For the schematic phase diagram in the

Discussion section and Table I we use the $T_N$ values derived from the signals $A_1$ that are best resolved and strongest yielding $T_N = 7$, 11 and 15 K for $x = 0.7$, 0.8 and 0.9, respectively.

While the transverse damping for $x = 0.8$ and 0.9 is strong below the magnetic ordering temperature due to static field distributions, the longitudinal damping affecting only the 1/3 "tail" is nearly vanishing as expected for static magnetic order. The behavior of the $x = 0.7$ sample is different: from the spectrum in Fig. 4c we can see that $G_z(t)$ is clearly decreasing below 1/3 at long times (measured up to 9 µs, not shown in the figure), i.e. there is considerable longitudinal damping due to fluctuating local fields. We interpret this as the partial presence of dynamic short-range order.

Further support for the suppression of magnetic order can be traced from the wTF experiments. In the paramagnetic regime the applied magnetic field leads to rotation signals with an asymmetry that corresponds to $A_{para} = A_{tot}$. When entering the magnetically ordered regime the randomly acting local fields in polycrystalline material will be much stronger than the weak applied field and lead to very strong damping of the oscillating signal from the applied field. This results in a reduction of the oscillating asymmetry $A_{para}$. By following the ratio $A_{para}(T)/A_{tot}$ we can trace the vanishing of the paramagnetic volume fraction upon lowering temperature. From Fig. 4e we can see that the transitions for $x = 0.9$, 0.8 and 0.7 are not sharp. The suppression of $A_{para}$ for $x = 0.9$ and 0.8 occurs close to, yet lower than those temperatures where spontaneous rotations could be resolved, which indicates inhomogenous order with contributions from short-range order. For $x = 0.7$ about 12% of volume stays paramagnetic even at 1.5 K consistent with our previous observation of strong longitudinal relaxation.

For all other concentrations $x = 0.1$-$0.6$ we could not trace a reduction of asymmetry under wTF down to 1.5 K (for $x = 0.5$ down to 19 mK as reported in ref. [32]). ZF spectra (Fig. 4f) show no indications of spontaneous muon spin rotation and longitudinal damping stays finite confirming a paramagnetic state down to the lowest temperatures measured for these compositions $x$. All ZF spectra reveal nearly identical depolarization except for the $x = 0.6$ sample, which has a clearly faster decay of asymmetry. Depolarization from nuclear dipolar fields is expected to be similar for all compositions $x$, therefore the reason has to lie in an additional dynamic contribution to relaxation. Whether this is caused by a close-lying critical point or the development of a small order parameter with an eventual spin-glass-like depolarization needs further experimental investigation.

### F. Density functional theory calculations

One of the main questions in understanding the origin of the spin-liquid-like state is the position of the materials in the $J_1$-$J_2$ phase diagram, i.e. whether they are in the Néel, columnar or highly frustrated region. We utilized *ab initio* density functional theory calculations to evaluate the relative stability of Néel and columnar orderings in $Sr_2Cu(Te_{1-x}W_x)O_6$. This is an extension of our previous DFT work on $Sr_2CuWO_6$[23]. It should be noted that the structural disorder on the Te/W *B''*-site is not included in the DFT calculations as they require a periodic system.

The Néel and columnar type orders are more stable than the ferromagnetic order for all compositions, see Supplemental Material[50]. This shows that both $J_1$ and $J_2$ are antiferromagnetic, and that there is magnetic frustration. In Fig. 5 we show the relative stability of Néel order compared to columnar order as a function of $x$ in $Sr_2Cu(Te_{1-x}W_x)O_6$. As expected, at $x = 0$ Néel order is clearly favored and at $x = 1$ columnar order is clearly favored. Notably, very little $W^{6+}$ is needed to destablize Néel order in $Sr_2Cu(Te_{1-x}W_x)O_6$: the crossover from Néel to columnar order occurs already at $x \approx 0.2$. A comparison of energies to a ferromagnetic state reveals that $W^{6+}$ destabilizes Néel order more strongly than it stabilizes columnar order, suggesting that the average $J_1$ is strongly reduced by $W^{6+}$ in addition to a weaker increase in average $J_2$. At $x = 0.5$, where we previously established a spin-liquid-like ground state, columnar order is favored indicating a dominant antiferromagnetic $J_2$ interaction ($J_2/J_1 > 0.5$). Thus, our calculations suggest that the origin of the spin-liquid-like state is not solely in the magnetic frustration of antiferromagnetic $J_1$ and $J_2$ interactions on the square lattice. In contrast to $x = 0$, the columnar order of $x = 1$ is quite stable towards $Te^{6+}$ substitution. Columnar order is expected to be very stable at least down to $x = 0.75$ in agreement with the μSR results.

Partial density of states (PDOS) were also investigated for $x = 0$, 0.5 and 1, see Supplemental Material[50]. The $W^{6+}$ $5d^0$ hybridization with O $2p$ was observed in $x = 1$ similar to previous DFT studies on $W^{6+}/Te^{6+}$ compounds[29,30]. This explains the strong 180° superexchange ($J_2$) via Cu-O-W-O-Cu favoring columnar order in the W-rich compounds. We also observe modest Te $5p$ – O $2p$ hybridization in $x = 0$, which is in agreement with a previous DFT study on $Sr_2CuTeO_6$[29]. In $x = 0.5$ we see both effects with significant W $5d^0$ – O $2p$ hybridization and modest Te $5p$ – O $2p$ hybridization.

## IV. DISCUSSION

A schematic phase diagram for $Sr_2Cu(Te_{1-x}W_x)O_6$ based on the μSR results and previous literature[24,25,32] is presented in Fig. 6. Néel order is significantly suppressed already at $x = 0.1$ in accordance with the DFT results. On the W-rich side, $T_N$ is gradually suppressed from 24 K at $x = 1$ to 7 K at $x = 0.7$. This is consistent with the DFT results, which predict that columnar antiferromagnetic order is very stable when $x = 0.75$-1. Of most importance is the wide frustrated region $x = 0.1$-0.6, where no magnetic order is observed at the lowest temperatures measured. These samples also have significant $T$-linear terms in the low-temperature specific heat as expected in a gapless QSL or a spin glass. We previously showed that $x = 0.5$ has a spin-liquid-like ground state with dynamic magnetism down to 19 mK. Our μSR evidence for a QSL-like state in the wider $x = 0.1$-0.6 region is not conclusive, as we could only measure down to 1.5-1.8 K. Consequently, we are not able to show a plateau in the muon spin relaxation rate expected in a QSL for these compositions and a spin glass ground state cannot be ruled out. The suppression of magnetic order in such a wide composition range does, however, indicate that the origin of the spin-liquid-like state is linked to both magnetic frustration and quenched disorder in the material.

The role of quenched disorder, inevitably present in a real material, is a non-trivial question in spin liquid physics[4]. Furukawa *et al.*[33] were able to induce a gapless spin liquid state in an organic triangular-lattice $Cu^{2+}$ salt by introducing disorder. Irradiation with x-rays created defects and disorder that drove the material from a magnetically ordered state into a spin liquid[33]. Disorder-induced spin liquid states were also observed in non-Kramers ion pyrochlores $Pr_2Zr_2O_7$ and $Tb_2Hf_2O_7$[34–36]. In comparison to these three compounds, the $Sr_2Cu(Te_{1-x}W_x)O_6$ series has even more quenched disorder due to being a solid solution.

Theoretical calculations by Kawamura and coworkers[37–40] predict the formation of disorder-induced QSL-like random singlet state on $S = 1/2$ triangular, kagome and honeycomb lattices. In the random singlet state, or valence bond glass, the quenched randomness in exchange interactions induces the formation of local spin singlets of varying strength with gapless excitations. In the honeycomb $J_1$-$J_2$ model, in which $J_1$ and $J_2$ compete similar to the square model, the random singlet state was stabilized in a wide parameter range in the presence of sufficient disorder[40]. Very recently, these theoretical calculations were extended to the square lattice[58]. For $J_2/J_1 = 0.5$, the ground state changes from a gapped spin liquid to a gapless random singlet state as quenched disorder is increased. The random singlet state is stabilized for a wide $J_2/J_1$ range for significant disorder. For $J_2/J_1 = 0.7$ in the columnar region, a spin glass state related to columnar magnetic order is stabilized for strong disorder. This stripe glass might be the ground state of $x = 0.6$. The experimental evidence previously presented for $x = 0.5$ is consistent with a random singlet state. The model is also consistent with a random singlet ground state for $x = 0.1$-$0.4$, as the spin glass is only stabilized close to columnar magnetic order in the phase diagram.

Other theoretical investigations of the effect of disorder on the square lattice are also known. In the unfrustrated case with only an antiferromagnetic $J_1$ interaction, Néel order is expected to be robust against bond disorder[38,59]. Richter[60] studied the effect of ferromagnetic inhomogeneities in the frustrated $J_1$-$J_2$ model. These were found to widen the quantum disordered region from $J_2/J_1 = 0.4$-$0.6$ to $0.15$-$0.6$, i.e. Néel order was destabilized but little effect was found on long-range order in the columnar region. This is quite different from the case in $Sr_2Cu(Te_{1-x}W_x)O_6$, where the frustrated region is widened towards the columnar region ($J_2/J_1 > 0.5$; $x > 0.2$).

## V. CONCLUSIONS

The entire solid solution series $Sr_2Cu(Te_{1-x}W_x)O_6$ ($0 \leq \underline{x} \leq 1$) was synthesized; the end members of the series being spin-1/2 square-lattice antiferromagnets with Néel ($x = 0$) and columnar ($x = 1$) order. Magnetic susceptibility measurements suggest a maximum in frustration occurs near $x \approx 0.5$, where a spin-liquid-like ground state was previously observed. In the specific heat, a $T$-linear term typical of gapless QSLs or spin glasses is observed for a wide composition range of $x = 0.1$-$0.7$. Muon spin rotation and relaxation measurements indicate, that long-range magnetic order is suppressed at least down to 1.5-1.8 K for a similar composition range of $x = 0.1$-$0.6$. These results suggest that the spin-liquid-like ground state of $x = 0.5$ could be stabilized for a wide composition range, although a spin glass ground state cannot be ruled out. The Néel magnetic order of $x = 0$ is destabilized with just 10% of Te-for-W substitution. In contrast, a gradual decrease in $T_N$ from 24 K to 7 K was found from $x = 1$ to 0.7 in the columnar side. Density functional theory calculations indicate that the compounds with $x > 0.2$ are all in the columnar region. Thus, magnetic frustration arising from competing $J_1$ and $J_2$ interactions alone cannot explain the suppression of magnetic order for $x = 0.1$-$0.6$ and the spin-liquid-like ground state of $x = 0.5$; a combination of frustration and the quenched disorder present in this solid solution is required.

**ACKNOWLEDGEMENTS**

The authors thank Dr. J.-C. Orain and Dr. J. Barker for technical assistance with the µSR measurements. Prof. H Kawamura is thanked for fruitful discussions on the effects of disorder. The authors wish to acknowledge CSC – IT Center for Science, Finland, for generous computational resources. O. Mustonen is grateful for an Aalto CHEM funded doctoral student position. S. Vasala is thankful for the support of the Brazilian funding agencies CNPq (grants no. 150503/2016-4 and 152331/2016-6) and FAPERJ (grant no. 202842/2016). F. Jochen Litterst, E. Sadrollahi and E. Baggio-Saitovitch are grateful for financial support by a joint DFG-FAPERJ project DFG Li- 244/12. In addition, E. Baggio-Saitovitch acknowledges support from FAPERJ through several grants including Emeritus Professor fellow and CNPq for BPA and corresponding grants.

Table I. Magnetic and thermodynamic properties of $Sr_2Cu(Te_{1-x}W_x)O_6$.

| $x$ | $0^{ab}$ | 0.1 | 0.2 | 0.3 | 0.4 | $0.5^a$ | 0.6 | 0.7 | 0.8 | 0.9 | $1^{ac}$ |
|---|---|---|---|---|---|---|---|---|---|---|---|
| $\mu_{eff}$ ($\mu_B$) | 1.87 | 1.8 | 1.84 | 1.82 | 1.85 | 1.87 | 1.84 | 1.88 | 1.87 | 1.9 | 1.9 |
| $\Theta_{cw}$ (K) | -80 | -64 | -66 | -62 | -63 | -71 | -80 | -96 | -109 | -138 | -165 |
| $T_{max}$ (K) | 74 | 68 | 62 | 58 | 52 | 52 | 52 | 56 | 64 | 72 | 86 |
| $\chi_{max}$ ($10^{-3}$ emu/mol) | 2.24 | 2.29 | 2.45 | 2.52 | 2.59 | 2.55 | 2.34 | 2.24 | 2.03 | 1.79 | 1.55 |
| $\gamma$ (mJ/molK$^2$) | 2.2 | 31.7 | 41.4 | 51.1 | 54.0 | 54.2 | 46.9 | 31.6 | 13.4 | 3.5 | 0.7 |
| $\theta_D$ (K) | 381 | 409 | 390 | 436 | 411 | 395 | 396 | 357 | 343 | 353 | 361 |
| $T_N$ (K) | 29 | <1.5 | <1.8 | <1.6 | <1.8 | <0.019 | <1.8 | 7 | 11 | 15 | 24 |
| $f = |\Theta_{cw}|/T_N$ | 2.8 | >42 | >36 | >34 | >35 | >3700 | >44 | 13.7 | 9.9 | 9.2 | 6.9 |

[a] Reference 32
[b] Reference 24
[c] Reference 21

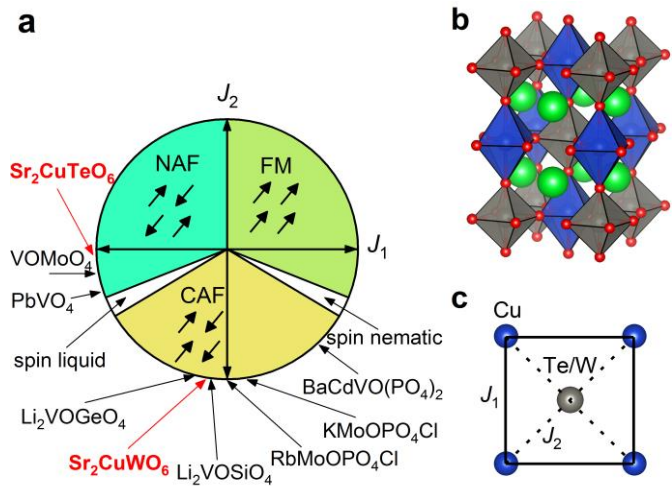

Fig 1. (a) Phase diagram of the spin-1/2 square-lattice Heisenberg model as a function of $J_2/J_1$, where $J_1$ is the nearest-neighbor and $J_2$ is the next-nearest-neighbor interaction[13,18,50]. Known compounds realizing the model are placed at their respective positions. The classical ground states of the model are ferromagnetic (FM) order, Néel antiferromagnetic (NAF) order and columnar antiferromagnetic (CAF) order. A quantum spin liquid state has been predicted for $J_2/J_1 \approx 0.4$-$0.6$ at the NAF-CAF boundary. Isostructural double perovskites $Sr_2CuTeO_6$ and $Sr_2CuWO_6$ are in the NAF and CAF regions, respectively. (b) The B-site ordered double perovskite structure of $Sr_2CuTeO_6$ and $Sr_2CuWO_6$[24,25]. (c) The square of $S = 1/2$ $Cu^{2+}$ cations in the $ab$ plane of $Sr_2CuTeO_6$ and $Sr_2CuWO_6$ with the view down the $c$-axis. The Te/W cation in the center of the $Cu^{2+}$ square determines whether $J_1$ or $J_2$ interaction dominates.

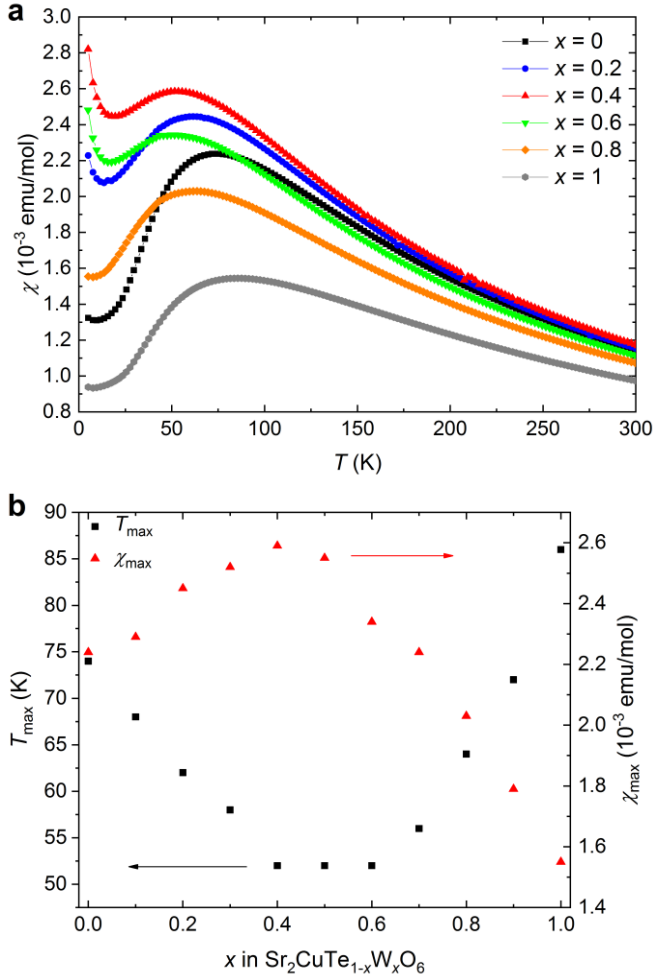

Fig 2. (a) Magnetic susceptibility of selected samples as a function of temperature. Zero-field cooled and field cooled data overlap and only the former is shown. (b) The position $T_{max}$ and height $\chi_{max}$ of the broad maximum in magnetic susceptibility as a function of $x$ in $Sr_2Cu(Te_{1-x}W_x)O_6$.

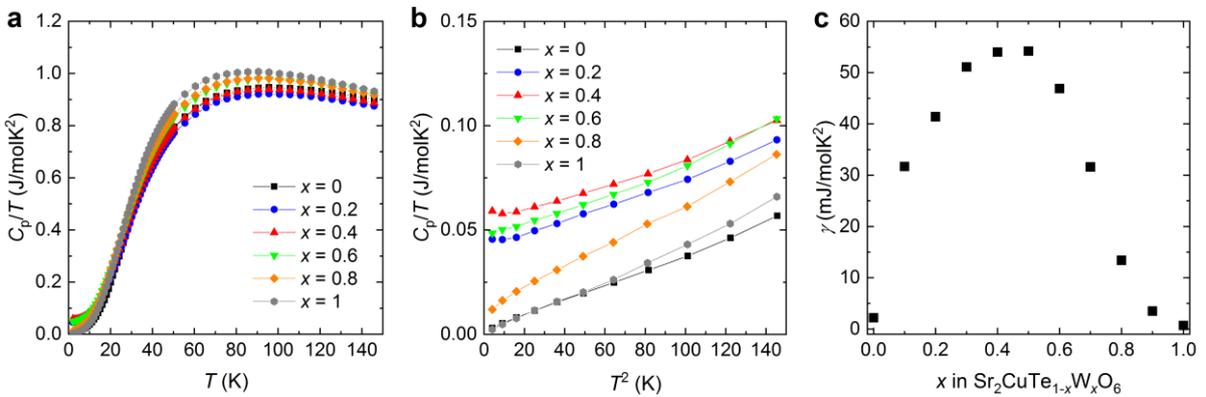

Fig 3. (a) Reduced specific heat of selected samples as a function of temperature. (b) Low-temperature $C_p$-$T^2$ plot showing the $T$-linear relationship of most samples. (c) $T$-linear term $\gamma$ of specific heat as a function of $x$ in $Sr_2Cu(Te_{1-x}W_x)O_6$.

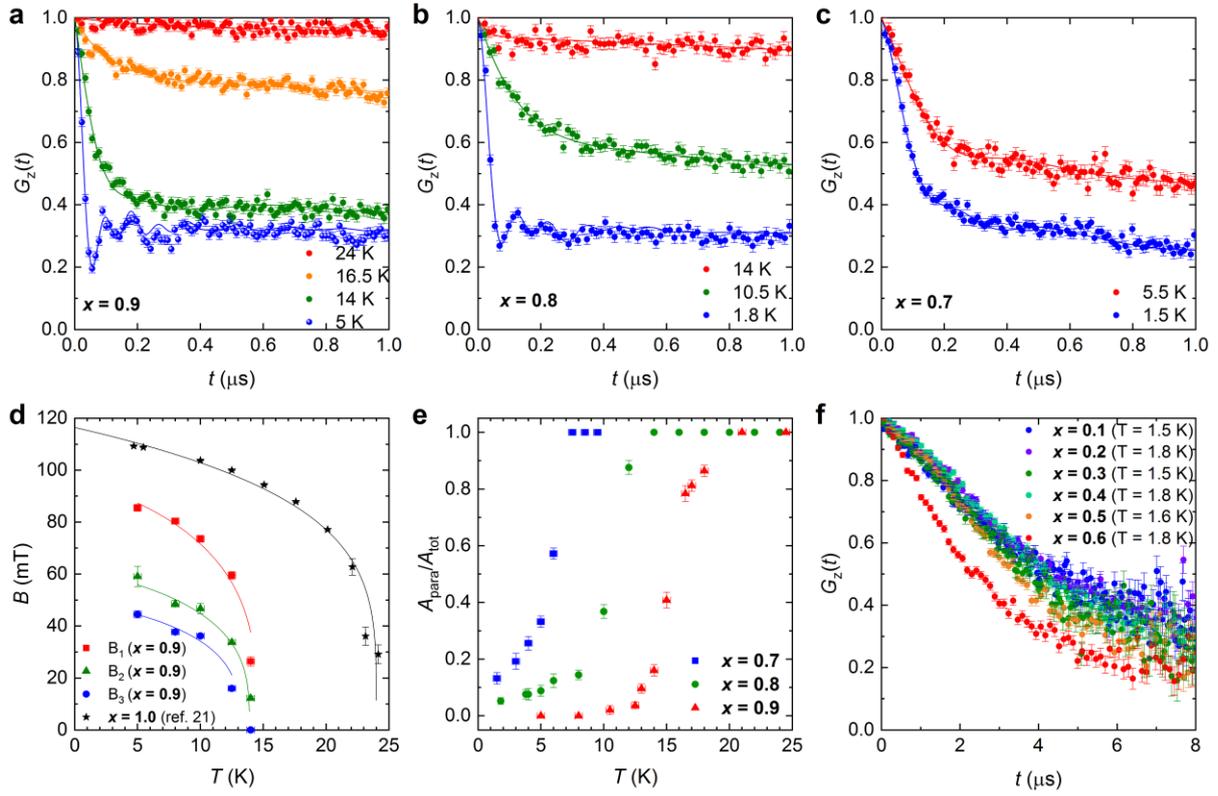

Fig 4. Zero-field spectra of $Sr_2Cu(Te_{1-x}W_x)O_6$ with (a) $x = 0.9$ and (b) $x = 0.8$ showing spontaneous rotation signals in the magnetically ordered state. (c) Zero-field spectra of $x = 0.7$ at 5.5 K and 1.5 K revealing only weak rotation signals and considerable dynamic damping. (d) Comparison of the temperature dependent local fields $B_i$ for $x = 0.9$ and $x = 1$ (data from ref. [21]). Drawn lines are interpolations using a standard approximation as described in the main text. (e) Temperature dependence of ratio between paramagnetic and total asymmetries measured in a weak transverse field of 5 mT for $x = 0.7, 0.8, 0.9$. (f) Zero-field spectra for $x = 0.1, 0.2, 0.3, 0.4, 0.5, 0.6$ at 1.5-1.8 K showing no indication of magnetic ordering or static magnetism.

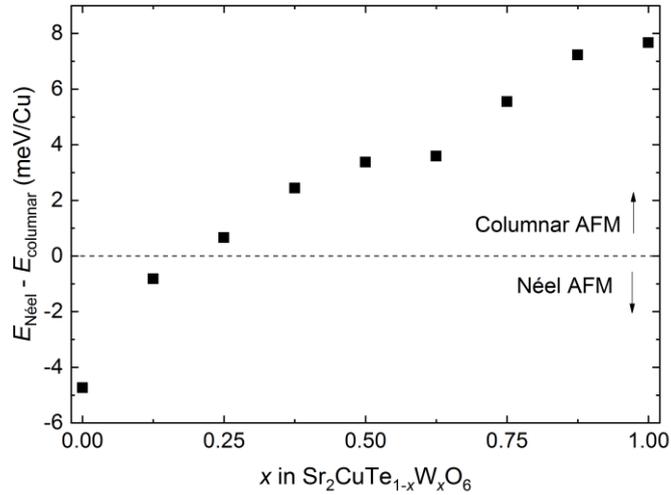

Fig 5. Energy difference between Néel and columnar antiferromagnetic order as a function of composition in $Sr_2Cu(Te_{1-x}W_x)O_6$ calculated using GGA+$U$.

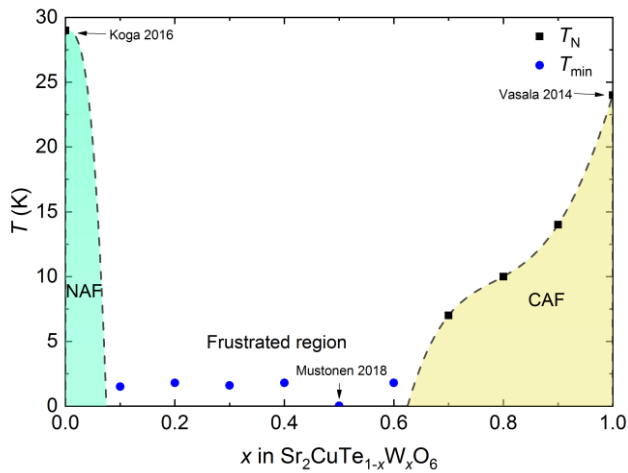

Fig 6. Schematic phase diagram of $Sr_2Cu(Te_{1-x}W_x)O_6$. The black squares represent measured Néel temperatures and the blue circles represent the lowest temperature measured for samples that remain entirely dynamic and do not order.

# Supplemental material for
# "Tuning the $S = 1/2$ square-lattice antiferromagnet $Sr_2Cu(Te_{1-x}W_x)O_6$ from Néel order to quantum disorder to columnar order"

O. Mustonen,[1] S. Vasala,[2] K. P. Schmidt,[3] E. Sadrollahi,[3] H. C. Walker,[4] I. Terasaki,[5] F. J. Litterst,[2,3] E. Baggio-Saitovitch,[2] M. Karppinen[1*]

[1]Department of Chemistry and Materials Science, Aalto University, FI-00076 Espoo, Finland

[2]Centro Brasileiro de Pesquisas Físicas (CBPF), Rua Dr Xavier Sigaud 150, Urca, Rio de Janeiro, 22290-180, Brazil

[3]Institut für Physik der Kondensierten Materie, Technische Universität Braunschweig, 38110 Braunschweig, Germany

[4]ISIS Neutron and Muon Source, Rutherford Appleton Laboratory, Chilton, Didcot, OX11 0QX, United Kingdom

[5]Department of Physics, Nagoya University, Nagoya 464-8602, Japan

Supplementary Table I. Rietveld refinement results for $Sr_2CuTe_{1-x}W_xO_6$. Space group $I4/m$ with atoms at W/Te (0, 0, 0), Cu (0, 0, 0.5), Sr (0, 0.5, 0.25), $O_1$ ($x$, $y$, 0), $O_2$ (0, 0, $z$).

| $x$ | 0[a] | 0.1 | 0.2 | 0.3 | 0.4 | 0.5[a] |
|---|---|---|---|---|---|---|
| $a$ (Å) | 5.43193(2) | 5.43156(2) | 5.43194(2) | 5.43200(2) | 5.43225(3) | 5.43192(3) |
| $c$ (Å) | 8.46750(3) | 8.45893(4) | 8.45248(5) | 8.44608(5) | 8.43994(5) | 8.43480(5) |
| $V$ (Å$^3$) | 249.842(1) | 249.555(2) | 249.399(2) | 249.215(2) | 249.057(2) | 248.875(2) |
| $O_1$ $x$ | 0.1932(7) | 0.2007(7) | 0.2016(7) | 0.2024(7) | 0.2031(8) | 0.2025(8) |
| $O_1$ $y$ | 0.2864(7) | 0.2879(6) | 0.2887(7) | 0.2873(7) | 0.2862(7) | 0.2866(8) |
| $O_2$ $z$ | 0.2223(6) | 0.2248(5) | 0.2258(6) | 0.2260(6) | 0.2260(6) | 0.2265(6) |
| Cu-O-Te/W (°) | 158.9(3) | 160.2(3) | 160.2(3) | 160.7(3) | 161.1(3) | 160.9(3) |
| $SrWO_4$ (%) | 0 | 0.1 | 0.4 | 0.4 | 0.5 | 0.6 |
| $R_p$ | 11.4 | 9.89 | 10.2 | 10.3 | 11.0 | 10.7 |
| $R_{wp}$ | 14.0 | 13.1 | 13.7 | 13.7 | 14.4 | 14.1 |
| $x$ | 0.6 | 0.7 | 0.8 | 0.9 | 1[a] | |
| $a$ (Å) | 5.43186(3) | 5.43136(2) | 5.43066(2) | 5.42996(1) | 5.42927(1) | |
| $c$ (Å) | 8.43021(5) | 8.42630(4) | 8.42277(3) | 8.42005(2) | 8.41682(2) | |
| $V$ (Å$^3$) | 248.734(2) | 248.573(2) | 248.405(2) | 248.261(1) | 248.103(1) | |
| $O_1$ $x$ | 0.2040(8) | 0.2029(8) | 0.2037(7) | 0.2037(6) | 0.2046(6) | |
| $O_1$ $y$ | 0.2869(8) | 0.2856(8) | 0.2858(7) | 0.2852(6) | 0.2845(5) | |
| $O_2$ $z$ | 0.2273(7) | 0.2259(6) | 0.2258(5) | 0.2260(4) | 0.2259(4) | |
| Cu-O-Te/W (°) | 161.2(3) | 161.2(3) | 161.3(3) | 161.5(2) | 161.8(2) | |
| $SrWO_4$ (%) | 0.4 | 0.5 | 0.8 | 0.9 | 0.5 | |
| $R_p$ | 11.1 | 10.5 | 8.86 | 7.81 | 7.2 | |
| $R_{wp}$ | 14.7 | 14.0 | 12.0 | 10.1 | 9.8 | |

[a] O. Mustonen *et al*. Nat. Commun. **9**, 1085 (2018).

Supplementary Table II. Energies obtained from the density functional theory calculations as a function of composition.

| $x$ in $Sr_2CuTe_{1-x}W_xO_6$ | 0 | 0.125 | 0.25 | 0.375 | 0.5 | 0.625 | 0.75 | 0.875 | 1 |
|---|---|---|---|---|---|---|---|---|---|
| $E_{FM}$ (meV/Cu) | 0 | 0 | 0 | 0 | 0 | 0 | 0 | 0 | 0 |
| $E_{Néel}$ (meV/Cu) | -8.03 | -2.99 | -2.11 | -2.06 | -2.65 | -1.27 | -0.50 | -2.77 | -3.36 |
| $E_{Columnar}$ (meV/Cu) | -3.30 | -2.17 | -2.77 | -4.50 | -6.02 | -4.86 | -6.05 | -9.99 | -11.03 |
| $E_{Néel}$ - $E_{Columnar}$ (meV/Cu) | -4.74 | -0.82 | 0.66 | 2.44 | 3.37 | 3.59 | 5.55 | 7.23 | 7.67 |

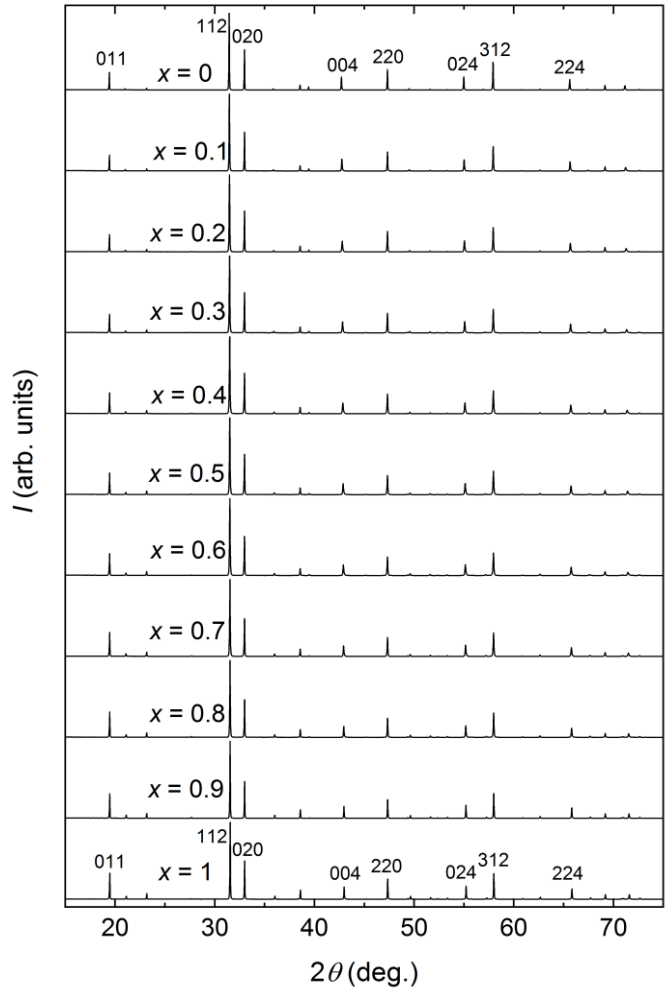

Supplementary Fig. 1. X-ray diffraction patterns of $Sr_2Cu(Te_{1-x}W_x)O_6$. Main reflections of the $I4/m$ space group are indexed. The main peak of the trace $SrWO_4$ impurity is too small to be visible in the figure.

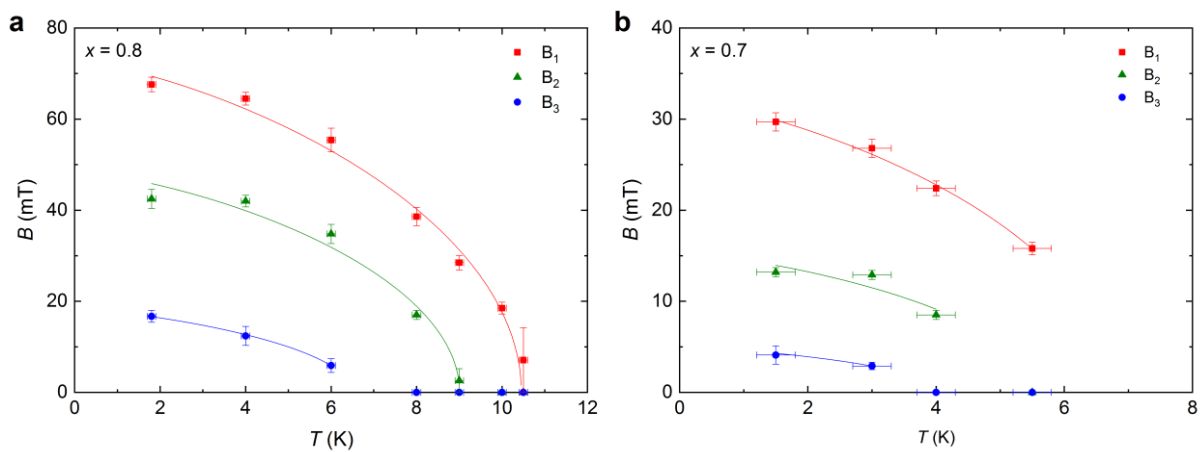

Supplementary Fig. 2. Temperature dependence of local fields $B_i$ for (a) $x = 0.8$ and (b) $x = 0.7$. Drawn lines are interpolations using a standard approximation as described in main text.

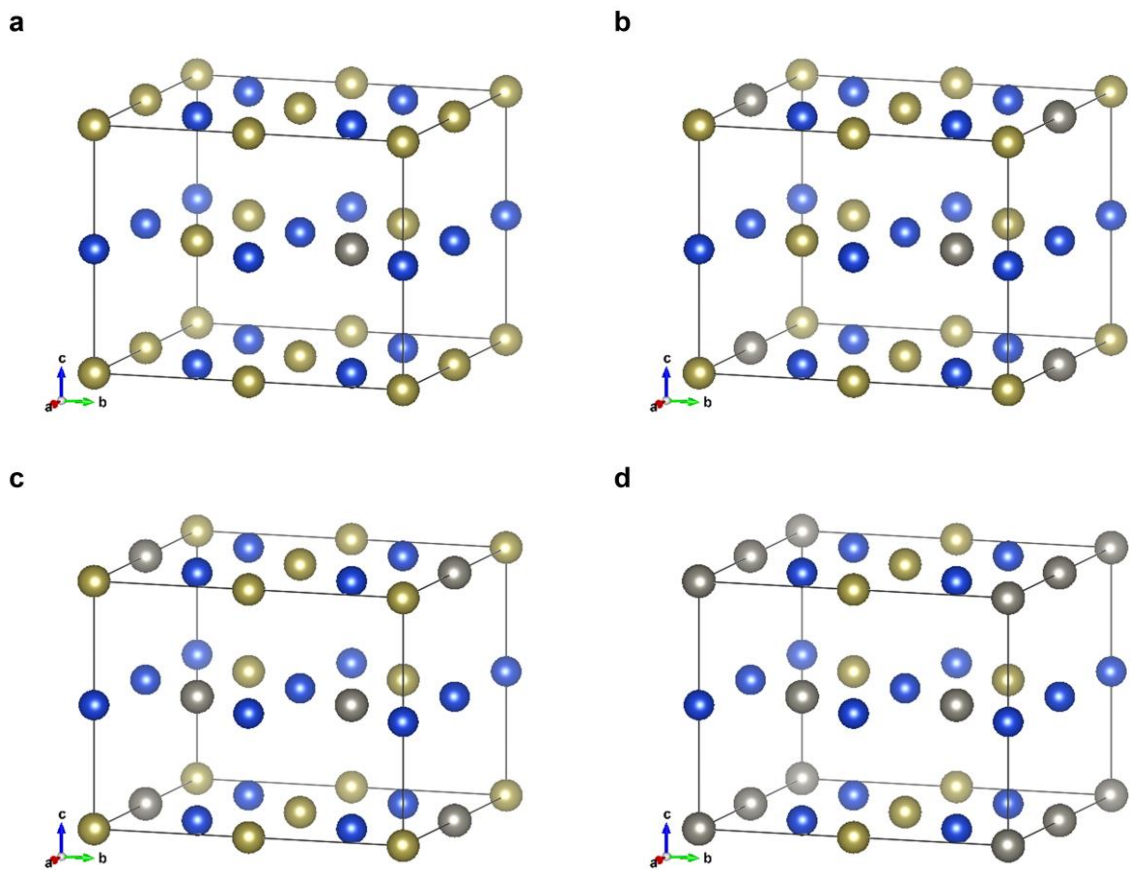

Supplementary Fig. 3. The 2 x 2 x1 supercells used for density functional theory calculations on the series. The dark gray (yellow) spheres represent the majority (minority) cation Te or W, and the blue spheres represent copper with strontium and oxygen omitted from the figure. a) $x = 0.125/0.875$ b) $x = 0.25/0.75$ c) $x = 0.375/0.625$ d) $x = 0.5$. For $x = 0.5$, each copper plaquette has an equal number of Te and W type plaquettes as neighbors to best describe the disorder in the system.

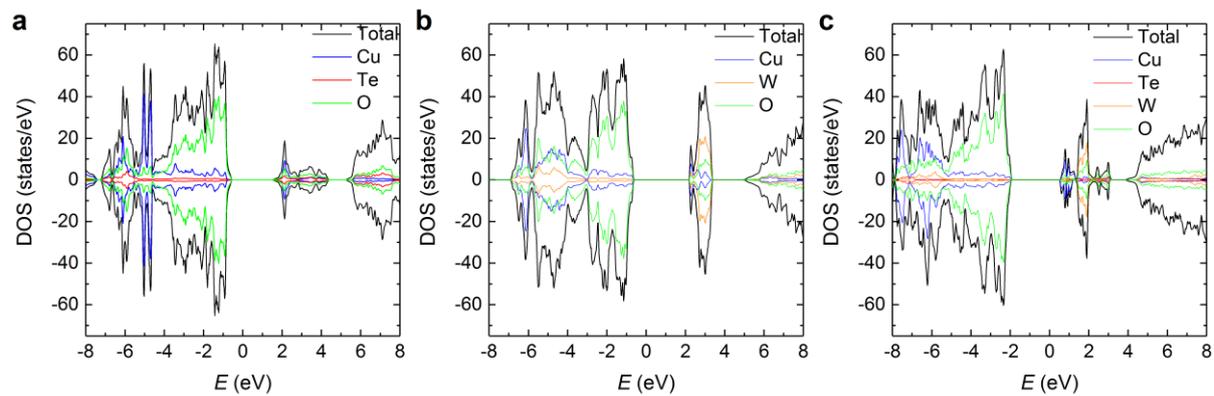

Supplementary Fig. 4. Total and partial density of states for a) $Sr_2CuTeO_6$, b) $Sr_2CuWO_6$ and c) $Sr_2CuTe_{0.5}W_{0.5}O_6$.